\documentclass[pra,twocolumn,superscriptaddress,amsmath,amssymb]{revtex4-2}
\usepackage{amsmath}
\usepackage{amsfonts}
\usepackage{amssymb}
\usepackage{graphicx}
\usepackage{color}
\usepackage{txfonts}
\usepackage{float}
\usepackage[titletoc]{appendix}
\usepackage[colorlinks={true}]{hyperref}
\usepackage{textcase}
\hypersetup{citecolor={blue}, filecolor={blue}, linkcolor={blue}, urlcolor={blue}}

\begin{document}
\title{Universal qutrit control in asymmetric-top molecules}
\author{Qian-Qian Hong}
\affiliation{Hunan Key Laboratory of Nanophotonics and Devices, Hunan Key Laboratory of Super-Microstructure and Ultrafast Process, School of Physics, Central South University,
Changsha, 410083, China}
\author{Zhi-Jian Zheng}
\affiliation{Hunan Key Laboratory of Nanophotonics and Devices, Hunan Key Laboratory of Super-Microstructure and Ultrafast Process, School of Physics, Central South University,
Changsha, 410083, China}

\author{Zhe-Jun Zhang}
\affiliation{Hunan Key Laboratory of Nanophotonics and Devices, Hunan Key Laboratory of Super-Microstructure and Ultrafast Process, School of Physics, Central South University,
Changsha, 410083, China}

\author{Xin-Xia Jian}
\affiliation{Hunan Key Laboratory of Nanophotonics and Devices, Hunan Key Laboratory of Super-Microstructure and Ultrafast Process, School of Physics, Central South University,
Changsha, 410083, China}

\author{Chuan-Cun Shu}
\email{cc.shu@csu.edu.cn}
\affiliation{Hunan Key Laboratory of Nanophotonics and Devices, Hunan Key Laboratory of Super-Microstructure and Ultrafast Process, School of Physics, Central South University,
Changsha, 410083, China}

\begin{abstract}
We present a theoretical framework for universal single-qutrit control in asymmetric-top molecules, advancing molecular quantum information processing. In this approach, the qutrit is encoded in three rotational eigenstates, with an auxiliary state providing independent phase control within the computational manifold. We explore an analytic protocol for arbitrary single-qutrit gates, combining directly addressable SU(2) rotations with auxiliary-state-mediated phase operations. To support this, we derive a multilevel pulse-area theorem that provides an explicit analytic mapping between gate parameters and control fields, enabling systematic design of high-fidelity microwave pulse sequences. Numerical simulations with 1,2-propanediol confirm the robustness of our approach, achieving Walsh–Hadamard gates with minimal leakage from the computational subspace. We further examine four SU(2) decomposition strategies and find that phase-error sensitivity depends on the decomposition sequence, while amplitude errors propagate along specific coherence pathways. Our results establish asymmetric-top molecules as a viable platform for qutrit-based quantum operations and offer an analytical method for precise quantum control of complex multilevel systems.
\end{abstract}

\maketitle
\section{Introduction}
High-dimensional quantum information processing can leverage the unique advantages of qudits, or $d$-level quantum systems, over conventional qubits \cite{2009NP_White,2019AQT_Cozzolino,2020FP_Kais,2020NJP_Sawant,2025RMD_Kiktenko}. By expanding the Hilbert space, qudits increase information density, simplify circuit design, and enable more complex entanglement. Qutrits ($d=3$) are particularly promising because they are sufficiently complex to exhibit high-dimensional phenomena while remaining experimentally accessible \cite{2021PRL_Morvan,2021PRX_Siddiqi,2022NC_Goss,2023PRL_Yu,2024NQI_Goss,2025PRL_Zeiher}. Achieving universal qutrit control requires the ability to perform arbitrary SU(3) operations, which can be decomposed into three SU(2) rotations on different two-level subspaces and a diagonal phase gate \cite{2012PRA_Vitanov,2014PLA_Dorai,2023PRApp_LeBlan,2023PRXQ_Tavernell}. The physical platform must therefore support direct, resonant coupling between all three level pairs to ensure full controllability. If some transitions are inaccessible, the system is restricted to a subgroup of SU(3), limiting the quantum algorithms that can be implemented.\\ \indent 
Many established physical platforms struggle to meet this requirement due to symmetry constraints. In superconducting circuits, anharmonicity often suppresses direct coupling between non-adjacent levels \cite{2020PRL_Ashhab,2021PRR_Lupascu,2023PRA_Murch}. In trapped ions and neutral atoms, dipole selection rules may prohibit certain transitions \cite{2022NP_Monz,2023NC_Martin,2026PRR_Ran}. Alternative approaches, such as indirect multi-photon processes or STIRAP-like transfers via auxiliary levels, can address these gaps but typically increase control complexity, sensitivity to decoherence, and operational times \cite{2016NC_Xu,2023PRApp_Roy}. Asymmetric-top molecules offer a natural solution to the SU(3) coupling requirement \cite{2020OE_Wu,2020PRApp_Wu,2022PRL_Li,2022PRA_shu,2022PRL_Lee,2023PRA_Zhang,2026CPC_shu}. Unlike linear or symmetric-top rotors, which often have at least one vanishing permanent electric dipole component due to symmetry, asymmetric tops generally possess three nonzero dipole components \cite{2019RMD_Sugny}. This enables one-photon electric-dipole coupling among any three rotational states allowed by selection rules, supporting the closed cyclic coupling necessary for universal control. Their anharmonic rotational spectra also allow precise spectral addressing of individual transitions using microwave fields \cite{2022CP_Koch,2024NC_Sandra,2025JPCL_Schnell}.\\ \indent 
We develop a universal control framework for single qutrits encoded in the rotational eigenstates of asymmetric-top molecules. This approach enables arbitrary single-qutrit gates through analytic construction of directly addressable SU(2) rotations and auxiliary-mediated phase operations. For control pulse-sequence design, we derive a multilevel pulse-area theorem that rigorously maps gate parameters to microwave control fields. Numerical simulations on 1,2-propanediol demonstrate high-fidelity Walsh–Hadamard gates with minimal computational leakage. Comparative analysis of four SU(3) decomposition sequences reveals how the order of SU(2) rotations affects operational robustness and error sensitivity. These results provide a strong theoretical foundation for realizing  single-qutrit control in molecular systems and high-dimensional quantum gates in multilevel architectures.
\\ \indent
The remainder of this paper is organized as follows. In Sec. \ref{sec2}, we present the theoretical framework for universal single-qutrit control in asymmetric-top molecules. Section \ref{sec3} provides numerical simulations and analyzes gate performance and robustness. Finally, Sec. \ref{sec4} summarizes the main findings and discusses future directions.

\begin{figure*}[htp]
\centering
\resizebox{0.9\textwidth}{!}{%
\includegraphics{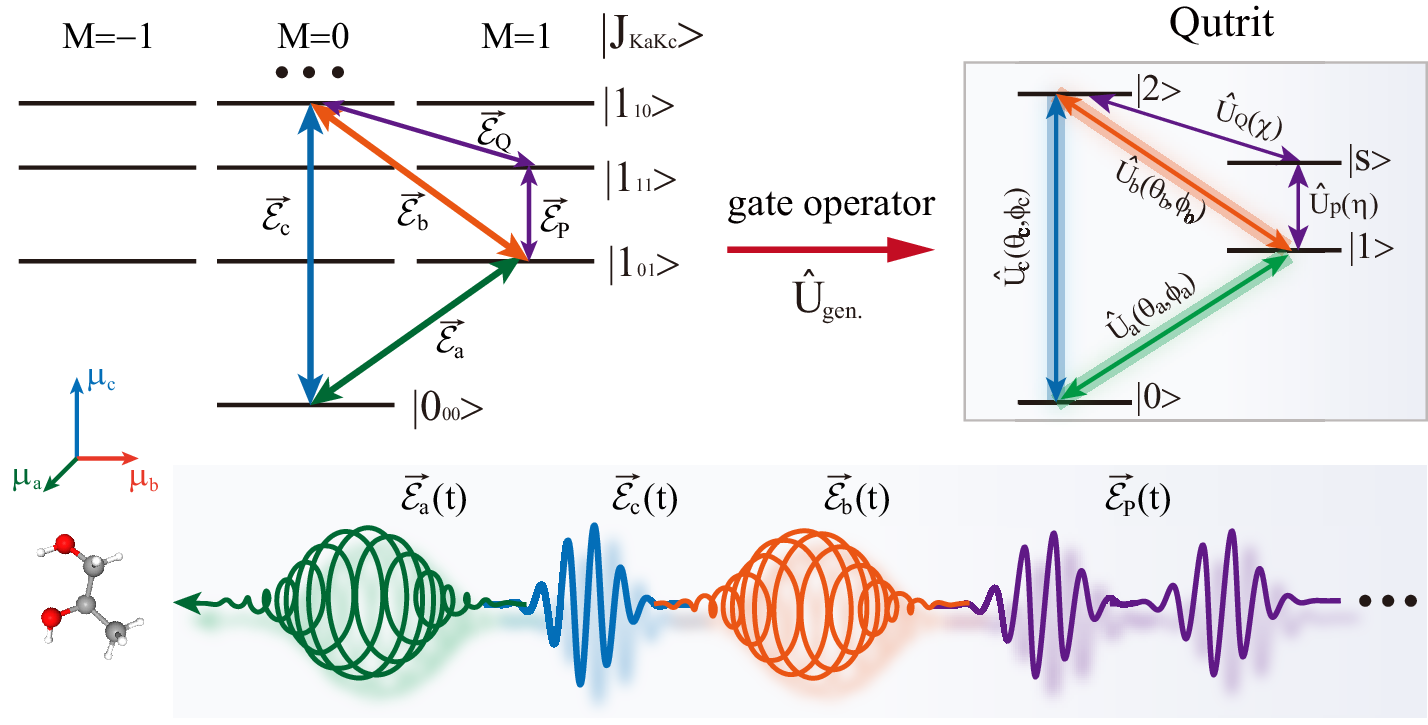} }\caption{Schematic of universal single-qutrit control in an asymmetric-top molecule. The qutrit is encoded in the rotational eigenstates $|0\rangle \equiv |0_{00},0\rangle$, $|1\rangle \equiv |1_{01},1\rangle$, and $|2\rangle \equiv |1_{10},0\rangle$. Pairwise transitions between qutrit states are driven by resonant microwave fields $\vec{\mathcal E}_{a,b,c}$, enabling implementation of the corresponding SU(2) operations
$\hat U_a$, $\hat U_b$, and $\hat U_c$. An auxiliary state $|S\rangle \equiv |1_{11},1\rangle$ allows for independent phase control of the qutrit manifold via two additional composite fields $\vec{\mathcal E}_{P,Q}$, enabling implementation of the diagonal phase operations $\hat U_P$ and $\hat U_Q$.} \label{fig1}
\end{figure*}

\section{THEORETICAL METHODS}\label{sec2}
In this section, we first present the theoretical foundations of asymmetric-top molecules driven by polarized microwave pulses, providing the basis for realizing molecular qutrits. We then show the decomposition of arbitrary single-qutrit gates into sequences of SU(2) rotations and a phase gate, followed by a systematic analysis of how different decomposition sequences respond to control errors.
\subsection{Microwave-driven asymmetric-top molecules}
We model asymmetric molecules as rigid rotors, focusing on microwave-driven rotational transitions in their electronic and vibrational ground states. The field-free rotational Hamiltonian is ($\hbar=1$) \cite{2019RMD_Sugny}
\begin{equation}
\hat{H}_0=A\hat{J}_a^2+B\hat{J}_b^2+C\hat{J}_c^2,
\end{equation}
where $A>B>C$ are rotational constants and $\hat{J}_{a,b,c}$ are angular momentum components along the molecule's principal axes.

For $A>B=C$ or $A=B>C$, the rotor reduces to a prolate or oblate symmetric top, with eigenstates $\left|JK_aM\right\rangle$ or $\left|JK_cM\right\rangle$. Here, $J$ is total angular momentum, $M$ and $K$ are its projections onto the space-fixed $Z$ axis and the molecular symmetry axis. In general asymmetric tops, $K$ is not a good quantum number, and eigenstates expand in the symmetric-top basis:
\begin{equation}
\left|J_{K_aK_c}M\right\rangle=\sum_{K}C_K^{J}\left|JKM\right\rangle,
\end{equation}
where  the expanding coefficients $C_K^J$ and energies $E_{J_{K_aK_c}}$ are calculated by solving $\hat H_0 \left|J_{K_aK_c}M\right\rangle=E_{J_{K_aK_c}} \left|J_{K_aK_c}M\right\rangle$, with $a\rightarrow z$, $b\rightarrow x$, $c\rightarrow y$. The pair $(K_a,K_c)$ indexes rotational levels within each $J$ manifold, corresponding to $|K|$ in the symmetric-top limits.\\ \indent 
In the electric-dipole approximation, the interaction between the molecules and the applied control fields $\vec{\mathcal{E}}(t)$ is described by \cite{1988Zare}
\begin{equation}
    \hat{H}_{\text{int}}(t)=-\vec{\hat{\mu}}\cdot\vec{\mathcal{E}}(t)=-\sum_{q=0,\pm1}(-1)^q\hat{\mu}_q^{\left(\mathrm{sf}\right)}\mathcal{E}_{-q}(t), 
\end{equation}
where $\hat{\mu}_0^{(\mathrm{sf})}=\hat{\mu}_Z$ and
$\hat{\mu}_{\pm1}^{(\mathrm{sf})}=\mp(\hat{\mu}_X\pm i\hat{\mu}_Y)/\sqrt{2}$ are the spherical components of the dipole operator in the space-fixed frame. These components relate to the molecule-fixed frame by $\hat{\mu}_q^{(\mathrm{sf})}=\sum_{q'=0,\pm1}D_{qq'}^{1*}\hat{\mu}_{q'}^{(\mathrm{mf})}$, where $D_{qq'}^{1}$ is the rank-1 Wigner D-matrix. The spherical components in the molecule-fixed frame are
$\hat{\mu}_0^{(\mathrm{mf})}=\hat{\mu}_a$ and
$\hat{\mu}_{\pm1}^{(\mathrm{mf})}=\mp(\hat{\mu}_b\pm i\hat{\mu}_c)/\sqrt{2}$, where $\mu_{a,b,c}$ are the permanent dipole moment components along the principal axes of the molecule-fixed frame. \\ \indent 
Transition matrix elements between asymmetric-top states are determined by Wigner D-matrix elements and can
be calculated using Wigner 3j-symbols:
\begin{equation}\label{tran1}
    \left\langle J'_{K_a'K_c'}M'\right| D_{qq'}^{1*}\left |J_{K_aK_c}M \right \rangle=\sum_{K,K'}\left(C_{K'}^{J'}\right)^*C_{K}^{J}\langle J'K'M'|D_{qq'}^{1*}|JKM\rangle,
\end{equation}
with
\begin{equation}\label{tran2}
\begin{aligned}
    \langle J'K'M'|D_{qq'}^{1*}|JKM\rangle&=\sqrt{2J+1}\sqrt{2J'+1}(-1)^{M'+K'+q-q'}\\&\times\begin{pmatrix} J & 1 & J' \\ M & -q & -M' \end{pmatrix}
\begin{pmatrix} J & 1 & J' \\ K & -q' & -K' \end{pmatrix}.
\end{aligned}
\end{equation}
The first $3j$ symbol gives $\Delta J=0,\pm1$ and $\Delta M=q$ (with $q = 0, \pm 1$ set by field polarization). The second governs selection rules for the molecule-fixed dipole components.
Electric-dipole transitions depend on the nonzero dipole-moment component: (1) $a$-type ($\mu_a\neq0$): $\Delta K_a=0$, $\Delta K_c=\pm1$; (2) $b$-type ($\mu_b\neq0$): $\Delta K_a=\pm1$, $\Delta K_c=\pm1$; (3) $c$-type ($\mu_c\neq0$): $\Delta K_a=\pm1$, $\Delta K_c=0$ \cite{1988Zare}. Constructing a cyclic ($\Delta$-type) three-state system, where each pair is coupled by a single-photon transition, requires all three transition types. Therefore, only asymmetric-top molecules with $\mu_a\neq0$, $\mu_b\neq0$, and $\mu_c\neq0$ enable such cyclic coupling. This feature creates a three-state rotational subspace in which any pair can be directly coupled by single-photon transitions, supporting qutrit encoding and universal single-qutrit control.\\ \indent 
\begin{table*}[t]
\caption{Parameters of the qutrit Walsh-Hadamard gate for different decomposition sequences.}
\label{tb1}
\begin{ruledtabular}
\begin{tabular}{c c c c c c c c c}
Sequence & $\theta_a$ & $\phi_a$ & $\theta_b$ & $\phi_b$ & $\theta_c$ & $\phi_c$ & $\eta$ & $\chi$ \\ \hline
$\hat{U}_{\mathrm{gen.}}^{(1)}=\hat{U}(\eta,\chi)\hat{U}_c\hat{U}_a\hat{U}_b$ & $\arcsin(1/\sqrt{3})$ & $7\pi/6$ & $\pi/4$ & $\pi/6$ & $\pi/4$ & $4\pi/3$ & $2\pi/3$ & $5\pi/6$ \\

$\hat{U}_{\mathrm{gen.}}^{(2)}=\hat{U}(\eta,\chi) \hat{U}_b \hat{U}_c \hat{U}_a$ & $\pi/4$ & $3\pi/2$ & $\pi/4$ & $0$ & $\arcsin(1/\sqrt{3})$ & $3\pi/2$ & $5\pi/6$ & $2\pi/3$ \\

$\hat{U}_{\mathrm{gen.}}^{(3)}=\hat{U}(\eta,\chi) \hat{U}_b \hat{U}_a \hat{U}_c$ & $\arcsin(1/\sqrt{3})$ & $3\pi/2$ & $\pi/4$ & $0$ & $\pi/4$ & $3\pi/2$ & $3\pi/2$ & $5\pi/6$ \\

$\hat{U}_{\mathrm{gen.}}^{(4)}=\hat{U}(\eta,\chi) \hat{U}_a \hat{U}_c \hat{U}_b$ & $\pi/4$ & $4\pi/3$ & $\pi/4$ & $11\pi/6$ & $\arcsin(1/\sqrt{3})$ & $7\pi/6$ & $5\pi/6$ & $3\pi/2$ \\
\end{tabular}
\end{ruledtabular}
\end{table*}
The time-dependent evolution of the molecule is characterized by the unitary operator $\hat{U}(t,t_0)$, which can be obtained by solving the time-dependent Schr\"odinger equation and has an exact solution in the interaction picture as \cite{2015PRA_Scully}
\begin{equation}\label{Ut}
    \hat{U}(t,t_0)
    =
    \hat{\mathcal{T}}
    \exp\!\left[
        -i
        \int_{t_0}^{t}
        e^{i\hat H_0 t'}
        \hat H_{\mathrm{int}}(t')
        e^{-i\hat H_0 t'}
        \, dt'
    \right],
\end{equation}
where $\hat{\mathcal{T}}$ denotes the time-ordering operator. The corresponding time-dependent state of the molecule is given by $|\psi(t)\rangle=\hat{U}(t,t_0)|\psi(t_0)\rangle$ with $|\psi(t_0)\rangle$ being the initial state. Equation (\ref{Ut}) provides an exact description of the driven rotational dynamics in the full Hilbert space.

\subsection{Universal single-qutrit gate decomposition}
A general SU(3) single-qutrit gate consists of three SU(2) rotations, each acting on a unique pair of qutrit states, and a diagonal phase gate. As illustrated in Fig.~\ref{fig1}, we encode the qutrit in three rotational states of an asymmetric-top molecule: $\left|0\right\rangle\equiv\left|{0_{00},0}\right\rangle$, $\left|1\right\rangle\equiv\left|{1_{01},1}\right\rangle$, and $\left|2\right\rangle\equiv\left|{1_{10},0}\right\rangle$, with eigenenergies \(E_0\equiv E_{0_{00}}\), \(E_1\equiv E_{1_{01}}\), and \(E_2\equiv E_{1_{10}}\). Control pulses $\vec{\mathcal{E}}_a(t)$, $\vec{\mathcal{E}}_b(t)$, and $\vec{\mathcal{E}}_c(t)$ selectively drive $a$-, $b$-, and $c$-type transitions. Applying a single control field confines the dynamics to the relevant two-level subspace, enabling an SU(2) rotation.

Under the first-order Magnus approximation \cite{2021PRA_shu,2023PRL_shu,2025PRR_Shu,2026PRA_shu}, the unitary operator controlled by each pulse in the qutrit basis $\left\{\left|0\right\rangle,\left|1\right\rangle,\left|2\right\rangle\right\}$ reads (see Appendix~\ref{ApP_1})
\begin{equation}
\begin{aligned}
\hat U_m(\theta_m,\phi_m)
=&\cos\theta_m\left(\left|i\right\rangle\left\langle i\right|+\left|j\right\rangle\left\langle j\right|\right)\\
&+i\sin\theta_m\left(e^{i\phi_m}\left|i\right\rangle\left\langle j\right|+\mathrm{H.c.}\right)+\left|k\right\rangle\left\langle k\right|,
\end{aligned}
\end{equation} where $i \neq j \neq k \in \{0,1,2\}$ and $m \in \{a,b,c\}$. The parameters $\theta_m$ and $\phi_m$ are set by the amplitude and phase of the complex pulse area $\mu_{ij}\int \mathcal{E}_m(t’)e^{-i\omega_{ij}t’}dt’$, where $\mu_{ij}$ is the transition dipole moment and $\omega_{ij}=|E_i-E_j|$ is the transition frequency for $\left|i\right\rangle\leftrightarrow\left|j\right\rangle$. It implies that each control pulse produces a coherent rotation between $|i\rangle$ and $|j\rangle$, while keeping $|k\rangle$ unaffected.

The phase gate employs an auxiliary state $\left|S\right\rangle\equiv\left|{1_{11},1}\right\rangle$ with energy $E_S\equiv E_{1_{11}}$ to imprint independent phases on $\left|1\right\rangle$ and $\left|2\right\rangle$ via loop transitions $\left|1\right\rangle\leftrightarrow\left|S\right\rangle$ and $\left|2\right\rangle\leftrightarrow\left|S\right\rangle$. Each loop is driven by a composite field: $\vec{\mathcal{E}}_P(t) = \vec{\mathcal{E}}_{P_1}(t) + \vec{\mathcal{E}}_{P_2}(t)$ for $\left|1\right\rangle\leftrightarrow\left|S\right\rangle$ and $\vec{\mathcal{E}}_Q(t) = \vec{\mathcal{E}}_{Q_1}(t) + \vec{\mathcal{E}}_{Q_2}(t)$ for $\left|2\right\rangle\leftrightarrow\left|S\right\rangle$. The corresponding complex pulse areas are defined as follows
\begin{equation}
\begin{aligned}
\theta_{P_j}e^{i\phi_{P_j}}&=\mu_{1S} \int \mathcal{E}_{P_j}(t’) e^{-i\omega_{1S} t’} \mathrm{d}t’,\ \\
\theta_{Q_j}e^{i\phi_{Q_j}}&=\mu_{2S} \int \mathcal{E}_{Q_j}(t’) e^{-i\omega_{2S} t’} \mathrm{d}t’,
\end{aligned}  \quad j=1,2
\end{equation}
$\mu_{1S}$ ($\mu_{2S}$) and $\omega_{1S}$ ($\omega_{2S}$) represent the respective transition dipole moments and transition frequencies. When the spectral amplitudes satisfy $\theta_{P_j} = \theta_P = \pi/2$ and $\theta_{Q_j} = \theta_Q = \pi/2$, the pulses implement diagonal phase operations $\hat{U}_P(\eta)$ and $\hat{U}_Q(\chi)$, respectively (see Appendix~\ref{ApP_1}). After the second subpulse, the population returns to the computational basis, and the target state acquires a geometric phase determined by the relative phases of the subpulses
\begin{equation}
\hat{U}(\eta,\chi) = \left|0\right\rangle\left\langle 0\right| + e^{i\eta}\left|1\right\rangle\left\langle 1\right| + e^{i\chi}\left|2\right\rangle\left\langle 2\right|,
\end{equation}where $\eta = \pi + \phi_{P_2} - \phi_{P_1}$ and $\chi = \pi - \phi_{Q_2} + \phi_{Q_1}$. The single-qutrit unitary can be decomposed by 
\begin{equation}\label{Ugen.}
\hat U_{\mathrm{gen.}}
=
\hat U(\eta,\chi)
\hat U_{m_1}(\theta_{m_1},\phi_{m_1})
\hat U_{m_2}(\theta_{m_2},\phi_{m_2})
\hat U_{m_3}(\theta_{m_3},\phi_{m_3}),
\end{equation}
where $m_1, m_2, m_3$ are distinct elements of $\{a, b, c\}$. Performing the three SU(2) operations results in six equivalent decompositions.
\subsection{Error analysis of decomposed single-qutrit gates}
Although different SU(2) decomposition sequences ideally implement the same target gate, a systematic comparison of their responses to control errors has not been conducted. To this end, we introduce relative parameter errors for each SU(2) operation: $\theta_m\rightarrow\theta_m(1+\xi)$ for amplitude and $\phi_m\rightarrow\phi_m(1+\zeta)$ for phase perturbations. Expanding the implemented gate $\hat U_{\mathrm{gen.}}(\alpha)$ ($\alpha\in\{\xi,\zeta\}$) to second order in $\alpha$ yields
\begin{equation}\label{Ualpha}
\hat U_{\mathrm{gen.}}(\alpha)=\hat U_{\mathrm{gen.}}(0)+\alpha\frac{\partial \hat U_{\mathrm{gen.}}}{\partial \alpha}\bigg|_{\alpha=0}+\frac{\alpha^2}{2}\frac{\partial^2 \hat U_{\mathrm{gen.}}}{\partial \alpha^2}\bigg|_{\alpha=0}+\mathcal{O}(\alpha^3).
\end{equation}
The average fidelity between the target $\hat U_{\mathrm{tar.}}$ and $\hat U_{\mathrm{gen.}}(\alpha)$ is (see Appendix~\ref{ApP_2})
\begin{equation}\label{gateerr}
\mathcal{F}_{\mathrm{gate}}(\alpha)=\frac{\left|\mathrm{Tr}\left[\hat U_{\mathrm{tar.}}^\dagger\hat U_{\mathrm{gen.}}(\alpha)\right]\right|^2+d}{d(d+1)}
=1-\frac{\alpha^2\mathcal{C}_\alpha^{\mathrm{gate}}}{d(d+1)}+\mathcal{O}(\alpha^3),
\end{equation}
where $\mathcal{C}_\alpha^{\mathrm{gate}} = d\,\mathrm{Tr}(\hat H_\alpha^2) - [\mathrm{Tr}(\hat H_\alpha)]^2$, $d=3$ for qutrit gates and $\hat{H}_\alpha = i\hat{U}_{\mathrm{tar.}}^\dagger\frac{\partial \hat U_{\mathrm{gen.}}}{\partial \alpha}\big|_{\alpha=0}$ is the Hermitian error generator. Since larger $\mathcal{C}_\alpha^{\mathrm{gate}}$ at fixed $\alpha$ implies greater fidelity loss, the gate's error sensitivity depends on the specific SU(2) decomposition chosen.\\ \indent
For a specific input state $|\psi(t_0)\rangle=|\psi_{\mathrm{in}}\rangle$, the ideal output state is $|\psi_{\mathrm{tar.}}\rangle=\hat{U}_{\mathrm{tar.}}|\psi(t_0)\rangle$, while the actual output state in the presence of parameter errors reads $|\psi_{\mathrm{out}}\rangle=\hat{U}_{\mathrm{gen.}}(\alpha)|\psi(t_0)\rangle$. The corresponding state fidelity is (see details in Appendix \ref{ApP_2})
\begin{equation}\label{stateerr}
\begin{aligned}
    \mathcal{F}_{\mathrm{state}}(\alpha)=\left|\left \langle \psi_{\mathrm{tar.}} | \psi_{\mathrm{out}} \right \rangle\right|^2= 1-\alpha^2(\langle\hat{H}_\alpha^2\rangle-\langle\hat{H}_\alpha\rangle^2)+\mathcal{O}(\alpha^3).
\end{aligned}
\end{equation}
The above expression shows that the leading-order reduction of the state fidelity is determined by the second-order coefficient $\mathcal{C}_\alpha^{\psi_{\mathrm{in}}}=\langle\hat{H}_\alpha^2\rangle-\langle\hat{H}_\alpha\rangle^2$. Consequently, $\mathcal{C}_\alpha^{\mathrm{gate}}$ and $\mathcal{C}_\alpha^{\psi_{\mathrm{in}}}$ provide quantitative measures for comparing the sensitivity of different decomposition sequences at both the gate level and the single-input-state level.
\subsection{Analytical design of the control pulse sequence}
The implementation of a single-qutrit gate relies on decomposing the target unitary into a sequence of elementary SU(2) operations acting on selected two-level subspaces, followed by a diagonal phase operation. In practice, each elementary operation is realized by a resonant control subpulse addressing a specific transition. The ordering of these subpulses, as specified by the decomposition, defines the overall control pulse sequence. Provided all control fields are phase-locked, any phase accumulated during pulse delays can be incorporated into the carrier phase, such that the envelope alone encodes the temporal delay \cite{2025PRA_Jian}. The resulting pulse sequence can be generally written as
\begin{equation}
\vec{\mathcal{E}}(t)
=\sum_n \vec{\mathcal{E}}_n(t)=\sum_n\epsilon_n f_n(t-T_n)\mathrm{Re}
\left[e^{-i(\omega_n t+\varphi_n)}\vec e_n\right],
\end{equation}
where $n \in \{a, b, c, P_1, P_2, Q_1, Q_2\}$. Here, $\epsilon_n$, $f_n(t)$, $T_n$, $\omega_n$, $\varphi_n$, and $\vec{e}_n$ denote the amplitude, envelope function, time delay, central frequency, phase, and polarization of the subpulse $\vec{\mathcal{E}}_n(t)$, respectively. For a subpulse $\vec{\mathcal{E}}_n(t)$ resonant with the transition $|i\rangle \leftrightarrow |j\rangle$, the complex pulse area determines the gate parameters $(\theta_n, \phi_n)$ for the corresponding SU(2) operation. This area is related to the Fourier transform of the control field: $\theta_n e^{i\phi_n} = \mu_{ij} A_n(\omega_{ij}) e^{i\varphi_n(\omega_{ij})}$, where
$A_n(\omega) e^{i\varphi_n(\omega)} = \int_{-\infty}^{\infty} \mathcal{E}_n(t') e^{-i\omega t'} dt'$, $A_n(\omega)$ and $\varphi_n(\omega)$ denote the spectral amplitude and phase, respectively.

Specifically, we consider each microwave pulse with a Gaussian spectral profile:
\begin{equation}
A_n(\omega)=A_n\exp\left[-\frac{(\omega-\omega_n)^2}{2\Delta\omega_n^2}\right]e^{-i(\omega-\omega_n) T_n},
\end{equation}
where $\Delta\omega_n$ is the bandwidth of the pulse. Setting $\omega_n = \omega_{ij}$ and performing the inverse Fourier transform yields the time-domain representation of each subpulse \cite{2023PCCP_shu,2025PRA_Fan,2026PRA_Yang}. The complete pulse sequence is then given by
\begin{equation}\label{Et}
\vec{\mathcal{E}}(t) =\sum_n\sqrt{\frac{2}{\pi}}\frac{\theta_n}{\tau_n|\mu_{ij}|}\exp{\left[-\frac{(t-T_n)^2}{2\tau_n^2}\right]}\mathrm{Re}\left[e^{-i(\omega_{ij} t+\varphi_n)}\vec e_n\right],
\end{equation}
where $\tau_n=1/\Delta\omega_n$ is the pulse duration and $\varphi_n=\phi_n-\arg(\mu_{ij})$. Different decompositions of the target gate correspond to different subpulse orderings, implemented through the relative delays $T_n$. The amplitudes and phases in Eq.~(\ref{Et}) are determined directly from the target gate parameters. This analytic mapping enables systematic and explicit design of pulse sequences for arbitrary single-qutrit gates.
\subsection{Average gate fidelity and state fidelity}
To evaluate the analytically designed pulse sequences, we numerically compute the exact rotational dynamics using the evolution operator $\hat{U}(t, t_0)$ [Eq.~(\ref{Ut})]. Following the control pulses ($t = t_f$), the molecular state is projected onto the computational qutrit basis using the projector $\hat{P}_q = |0\rangle\langle 0| + |1\rangle\langle 1| + |2\rangle\langle 2|$, resulting in the effective operation $\hat{M} = \hat{P}_q \hat{U}(t_f, t_0) \hat{P}_q$. The average gate fidelity relative to the target gate $\hat{U}_{\mathrm{tar.}}$ is \cite{2002PLA_Nielsen}
\begin{equation}\label{Figate}
\mathcal{F}(\hat{U}_{\mathrm{tar.}}) = \frac{\mathrm{Tr}(\hat{M}^\dagger \hat{M}) + |\mathrm{Tr}(\hat{U}_{\mathrm{tar.}}^\dagger \hat{M})|^2}{d(d+1)}, \quad d = 3,
\end{equation}
The state fidelity for an input $|\psi(t_0)\rangle = |\psi_{\mathrm{in}}\rangle$ is
\begin{equation}\label{Fidtate}
\mathcal{F}(\psi_{\mathrm{in}}) = \left| \langle \psi_{\mathrm{tar.}} | \hat{M} | \psi(t_0) \rangle \right|^2,
\end{equation}
where $|\psi_{\mathrm{tar.}}\rangle = \hat{U}_{\mathrm{tar.}} |\psi(t_0)\rangle$.
Population leakage to noncomputational states reduces both the average gate and state fidelities in Eqs.~(\ref{Figate}) and (\ref{Fidtate}).
Within the computational subspace, the qutrit state is described by the density matrix $\hat{\rho}(t) = |\psi(t)\rangle \langle \psi(t)|$.
Any qutrit density matrix admits an expansion in the Gell-Mann basis $\{ \hat{\lambda}_k \}_{k=1}^8$ (see Appendix~\ref{Appl}) \cite{2000OC_Milburn}:
\begin{equation}\label{rho}
\hat{\rho}(t) = \frac{1}{3} \hat{I} + \frac{1}{2} \sum_{k=1}^8 s_k(t) \hat{\lambda}_k,
\end{equation}
where $\hat{I}$ is the $3 \times 3$ identity, and $s_k(t) = \mathrm{Tr}[\hat{\rho}(t) \hat{\lambda}_k] = \langle \hat{\lambda}_k \rangle$ are the generalized Bloch vector components, fully characterizing the qutrit state. These definitions provide quantitative metrics for assessing the fidelity of gate operations and state evolution within the molecular qutrit subspace.
\begin{figure*}[t]
\centering
\resizebox{0.9\textwidth}{!}{%
\includegraphics{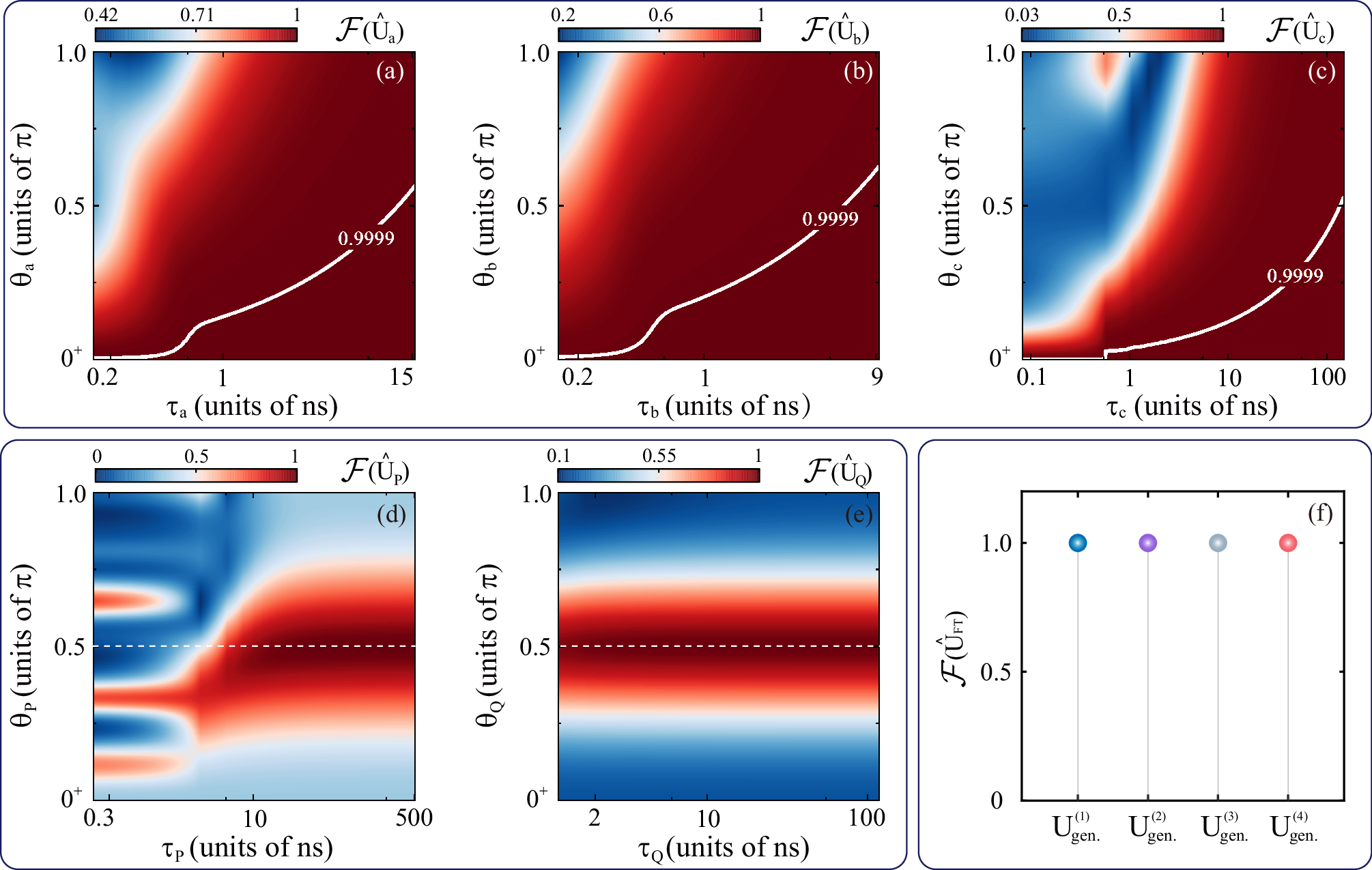} }\caption{Average gate fidelities for elementary operations and the target Walsh–Hadamard qutrit gate in 1,2-propanediol. (a-c) Fidelities $\mathcal{F}(\hat{U}_{a,b,c})$ of SU(2) rotations and (d,e) fidelities $\mathcal{F}(\hat{U}_{P,Q})$ of the auxiliary phase operations as functions of the pulse area $\theta_n$ and pulse duration $\tau_n$. (f) Fidelities $\mathcal{F}(\hat U_{\rm FT})$ of the Walsh–Hadamard gate for four admissible decomposition sequences $\hat U_{\rm gen.}^{(1–4)}$, obtained with pulse areas $\theta_{a,b,c}$ from Table \ref{tb1} and $\theta_{P,Q}=\pi/2$, and with pulse durations fixed at $\tau_a=6.3$ ns, $\tau_b=7$ ns, $\tau_c=163.8$ ns, $\tau_P=446.1$ ns, and $\tau_Q=94.3$ ns.} 
\label{fig2}
\end{figure*}

\section{Results and Discussion}\label{sec3}
As a representative example of a single-qutrit gate that encompasses all eight independent parameters of SU(3), we consider the Walsh-Hadamard gate, also known as the single-qutrit Fourier transform \cite{2003PRA_Klimov}:
\begin{equation}
\hat{U}_{\mathrm{FT}} = \frac{1}{\sqrt{3}} \begin{pmatrix}
1 & 1 & 1 \\
1 & e^{i2\pi/3} & e^{-i2\pi/3} \\
1 & e^{-i2\pi/3} & e^{i2\pi/3}
\end{pmatrix}.
\end{equation}
By combining $\hat{U}_{\mathrm{FT}}$ with Eq.~(\ref{Ugen.}), we determine gate parameters for elementary operations using different orderings of three SU(2) rotations. Among the six possible orderings, only four yield real-valued rotation angles and are physically feasible. Table~\ref{tb1} lists the relevant parameters, with detailed derivations in Appendix~\ref{para}. We conduct numerical simulations for the asymmetric-top molecule 1,2-propanediol, which has rotational constants $A = 8572.05$ MHz, $B = 3640.10$ MHz, $C = 2790.96$ MHz, and permanent dipole moment components $\mu_a = 1.2$ D, $\mu_b = 1.9$ D, and $\mu_c = 0.36$ D \cite{2013Nature_Patterson}.
\begin{figure}[t]
\centering
\resizebox{0.47\textwidth}{!}{%
\includegraphics{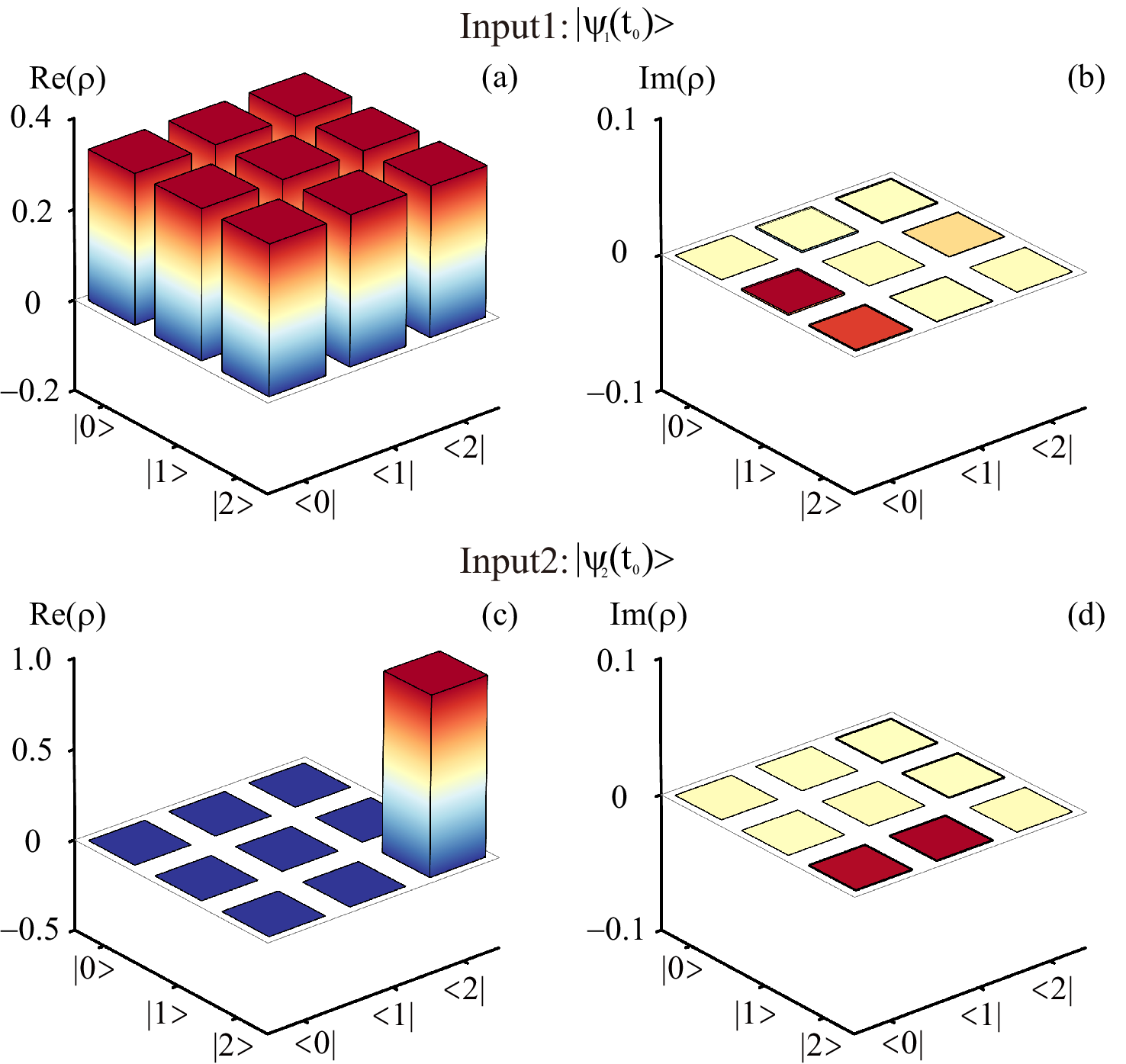} }\caption{Real (a,c) and imaginary (b,d) parts of the output density matrices after applying the Walsh–Hadamard gate to two representative input states. Panels (a) and (b) correspond to the basis input state $|\psi_1(t_0)\rangle=|0\rangle$, while panels (c) and (d) correspond to the coherent superposition state $|\psi_2(t_0)\rangle=1/\sqrt{3}(|0\rangle+e^{i2\pi/3}|1\rangle+e^{-i2\pi/3}|2\rangle)$. The amplitudes and phases of the pulse sequence are chosen according to the first decomposition listed in Table \ref{tb1}. } 
\label{fig3}
\end{figure}
\subsection{Implementation of the qutrit Walsh-Hadamard gate}
We evaluate each subpulse's ability to perform its assigned elementary operation. Using the pulse phases specified in Table~\ref{tb1}, Figs.~\ref{fig2}(a)–(e) show the average fidelities for the three SU(2) rotations, $\mathcal{F}(\hat{U}_{a,b,c})$, and two auxiliary phase operations, $\mathcal{F}(\hat{U}_{P,Q})$  as functions of pulse area $\theta_n$ and duration $\tau_n$. The analytically designed fields in  Eq.~(\ref{Et}) are used for the simulations. Fidelities are highly sensitive to both parameters. For a fixed pulse area, fidelity increases with longer duration, exceeding 0.9999 for sufficiently long pulses [Figs.~\ref{fig2}(a)–(c)], but decreases for short pulses. This results from the inverse relationship between pulse duration and spectral bandwidth: longer pulses produce narrower spectra, thereby suppressing excitation to other rotational states and reducing leakage from the computational subspace. For a fixed pulse duration, increasing the pulse area (stronger driving) increases leakage. To maintain spectral selectivity and achieve high-fidelity gates, stronger driving requires longer pulses.\\ \indent 
The fidelities for the phase operations in Figs.~\ref{fig2}(d) and (e) indicate that independent phase control depends on specific pulse areas. High-fidelity phase control is achieved when $\theta_P=\theta_Q=\pi/2$, consistent with analytic predictions. We therefore set $\tau_a=6.3$ ns, $\tau_b=7$ ns, $\tau_c=163.8$ ns, $\tau_P=446.1$ ns, and $\tau_Q=94.3$ ns to ensure all elementary operations achieve high fidelity. Pulse areas and phases are selected as in Table~\ref{tb1}. The resulting fidelities of the Walsh–Hadamard gate for the four valid decompositions are shown in Fig.~\ref{fig2}(f). In all cases, gate fidelity exceeds 0.9999, confirming that the analytic scheme enables high-fidelity qutrit gate implementation in the molecular system.\\ \indent
We evaluate the Walsh–Hadamard gate by analyzing output states for representative initial conditions using the first admissible decomposition sequence $\hat{U}_{\rm gen.}^{(1)}$. For the computational basis input $|\psi_1(t_0)\rangle = |0\rangle$, the resulting density matrix [Figs.~\ref{fig3}(a), (b)] closely matches the ideal Walsh–Hadamard output. The real part shows uniform population across all three basis states, and the imaginary part remains negligible, confirming the expected output structure.
\begin{figure}[t]
\centering
\resizebox{0.48\textwidth}{!}{%
\includegraphics{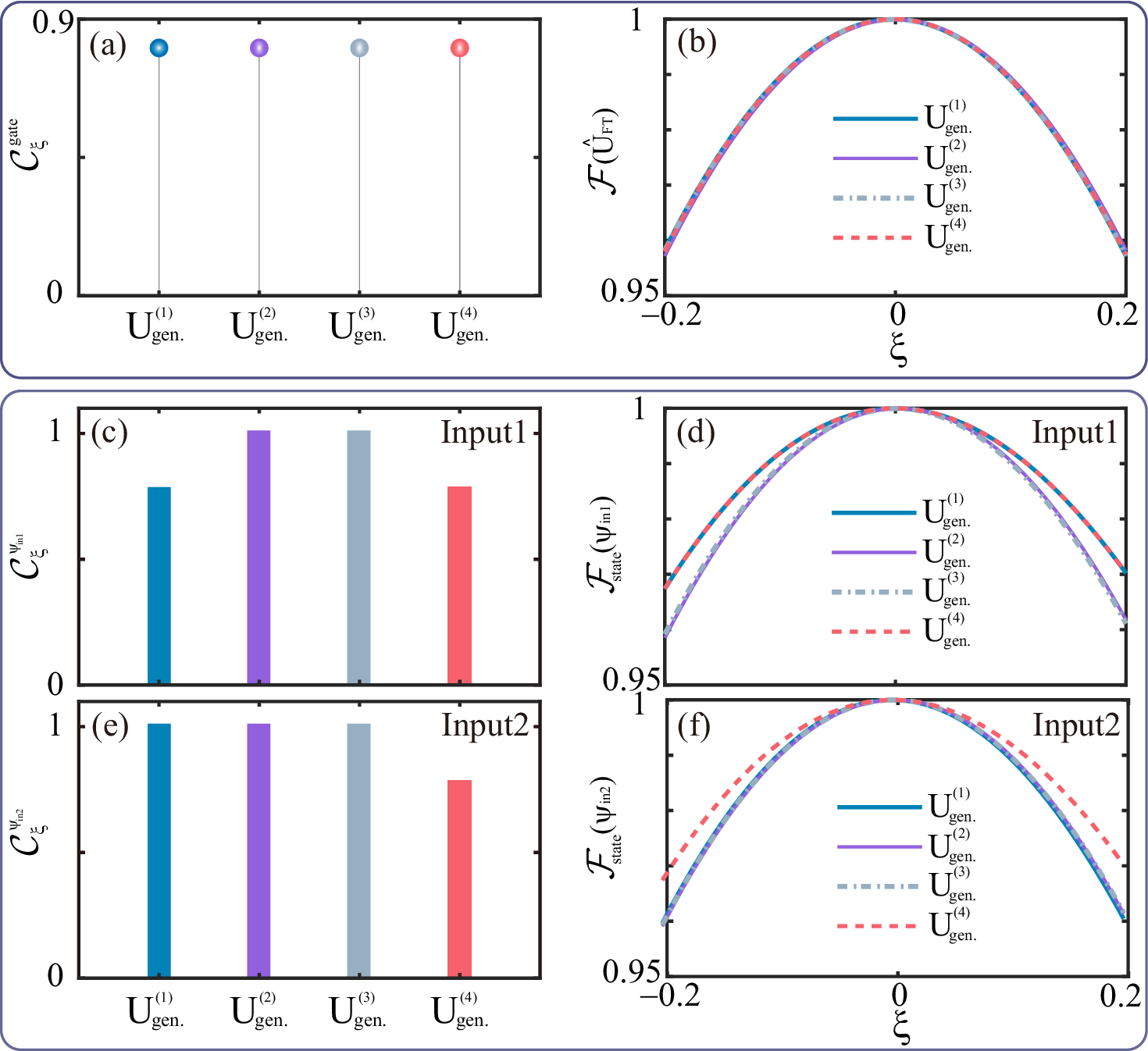} }\caption{(a) Analytical gate error coefficient $C_{\xi}^{\rm gate}=d\,\mathrm{Tr}(\hat H_\alpha^2)-\mathrm{Tr}(\hat H_\alpha)^2$ [Eq.~(\ref{gateerr})] and (c,e) analytical state error coefficients $C_{\xi}^{\psi_{\rm in}}=\langle \hat H_\alpha^2\rangle-\langle \hat H_\alpha\rangle^2$ [Eq.~(\ref{stateerr})] for input states $|\psi_1(t_0)\rangle=|0\rangle$ and $|\psi_2(t_0)\rangle=\frac{1}{\sqrt{3}}(|0\rangle+e^{i2\pi/3}|1\rangle+e^{-i2\pi/3}|2\rangle)$. (b) Average gate fidelity $\mathcal{F}(\hat U_{\rm FT})$ [Eq. (\ref{Figate})] and (d,f) corresponding state fidelities $\mathcal{F}_{\rm state}(\psi_{\rm in})$ [Eq. (\ref{Fidtate})] as functions of the amplitude error $\xi$, defined by $\theta_{a,b,c}\rightarrow(1+\xi)\theta_{a,b,c}$, for four decomposition sequences $\hat U_{\rm gen.}^{(1\text{-}4)}$. Other pulse parameters are the same as in Fig.~\ref{fig2}(f).} 
\label{fig4}
\end{figure}
To assess phase transformation, we use the coherent superposition input 
\begin{equation}
   |\psi_2(t_0)\rangle = \frac{1}{\sqrt{3}} \left(|0\rangle + e^{i2\pi/3}|1\rangle + e^{-i2\pi/3}|2\rangle \right).
\end{equation}
The resulting output density matrix [Figs.~\ref{fig3}(c), (d)] is dominated by a single diagonal element for $|2\rangle$, showing that the superposition is coherently mapped to a specific computational basis state. This outcome demonstrates constructive and destructive interference of input phases, consistent with the ideal Walsh–Hadamard transformation. These results confirm that the analytically designed pulse sequences accurately implement the target qutrit gate.
\begin{figure}[t]
\centering
\resizebox{0.5\textwidth}{!}{%
\includegraphics{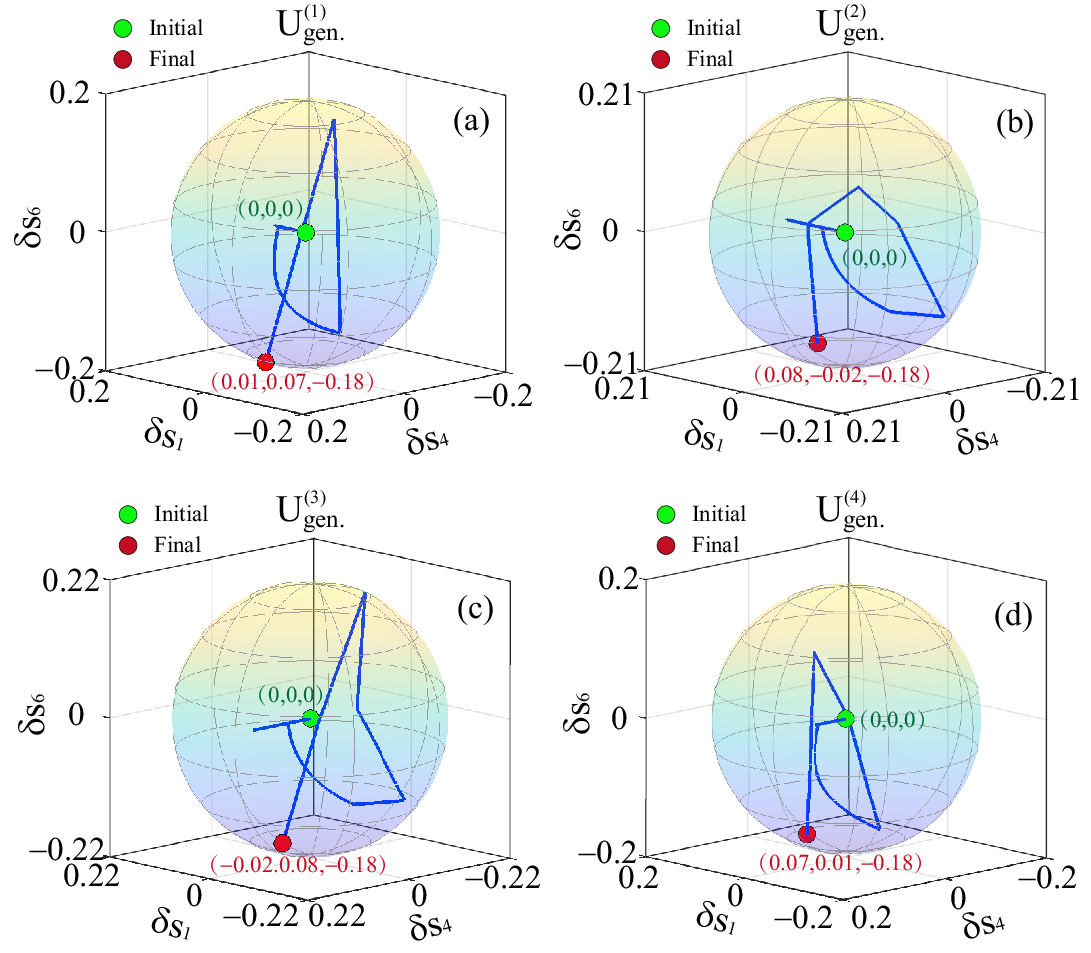} }\caption{Error dynamics in the generalized Bloch-vector representation under amplitude error. Time evolution of the deviation $\delta \mathbf{s}(t)=\mathbf{s}(\xi=-0.2,t)-\mathbf{s}(\xi=0,t)$ projected onto the subspace $(\delta s_1,\delta s_4,\delta s_6)$ for input state $|\psi_1(t_0)\rangle=|0\rangle$. Panels (a–d) correspond to the four decomposition sequences $\hat U_{\rm gen.}^{(1–4)}$.} 
\label{fig5}
\end{figure}
\subsection{Sensitivity to amplitude errors}

We evaluate the sensitivity of four SU(2) decomposition sequences to control errors. Equations~(\ref{gateerr}) and (\ref{stateerr}) indicate that reductions in average gate fidelity and target-state fidelity are determined by analytical error coefficients; larger coefficients reflect greater sensitivity to parameter perturbations. In the molecular qutrit system, we simulate amplitude errors by applying relative errors to the rotation angles, \(\theta_{a,b,c} \rightarrow (1+\xi)\theta_{a,b,c}\).

Figure~\ref{fig4} shows the analytical error coefficients
$\mathcal{C}_{\xi}^{\rm gate}$ and $\mathcal{C}_{\xi}^{\rm \psi_{in}}$, together with the
corresponding fidelities obtained from molecular simulations as functions of the error parameter $\xi$.
Figures~\ref{fig4}(a) and (b) indicate that all four decomposition sequences have identical analytical gate-error coefficients. The average gate fidelities exhibit the same dependence on amplitude error, confirming that target gate sensitivity to amplitude perturbations is independent of SU(2) ordering, in line with analytical predictions.

For fixed input states, target-state fidelity sensitivity depends on the decomposition sequence. For input state $|\psi_1(t_0)\rangle = |0\rangle$ [Figs.~\ref{fig4}(c) and (d)], the first and fourth decompositions share the same analytical coefficient and dependence on $\xi$, while the second and third decompositions form a separate pair with matching sensitivity.
For input state $|\psi_2(t_0)\rangle = \frac{1}{\sqrt{3}}(|0\rangle + e^{i2\pi/3}|1\rangle + e^{-i2\pi/3}|2\rangle)$ [Figs.~\ref{fig4}(e) and (f)], the first three decompositions have identical analytical coefficients and the same state-fidelity dependence on $\xi$, whereas the fourth decomposition shows distinctly different sensitivity.

In all cases, numerical results closely match the trends predicted by the analytical error coefficients: larger coefficients correspond to greater fidelity degradation under amplitude perturbations. Although the four decomposition sequences are equivalent at the target gate level, their robustness to amplitude errors can differ significantly for specific input states. Analytical error coefficients provide a quantitative basis for selecting decomposition sequences optimized for particular input states in qutrit control.

To further illustrate error accumulation during gate operation, we analyze the error dynamics in the generalized Bloch-vector representation for the input state $|\psi_1(t_0)\rangle=|0\rangle$. According to Eq.~(\ref{rho}), the ideal Walsh-Hadamard output has only three nonzero Bloch components, $s_1$, $s_4$, and $s_6$. We set $\xi=-0.2$ and visualize the deviations
$\delta s_k(t)=s_k(\xi,t)-s_k(0,t)$ in the reduced subspace
$(\delta s_1,\delta s_4,\delta s_6)$, as shown in Fig.~\ref{fig5}. The trajectories start from the zero-deviation point and follow sequence-dependent paths. Although the transient dynamics differ, a common feature emerges: the error is mainly concentrated in the $\delta s_6$ component, whereas $\delta s_1$ and $\delta s_4$ remain small.

In the Gell-Mann representation (see Appendix~\ref{Appl}), $\lambda_6$ corresponds to the real part of coherence between qutrit states $|1\rangle$ and $|2\rangle$. The dominance of $\delta s_6$ indicates that amplitude errors primarily appear as deviations in this coherence channel, leading to preferential error accumulation during multistep evolution. Although errors occur throughout the SU(2) sequence, their overall effect is concentrated mainly in a single channel.

\subsection{Sensitivity to phase errors}
\begin{figure}[t]
\centering
\resizebox{0.48\textwidth}{!}{%
\includegraphics{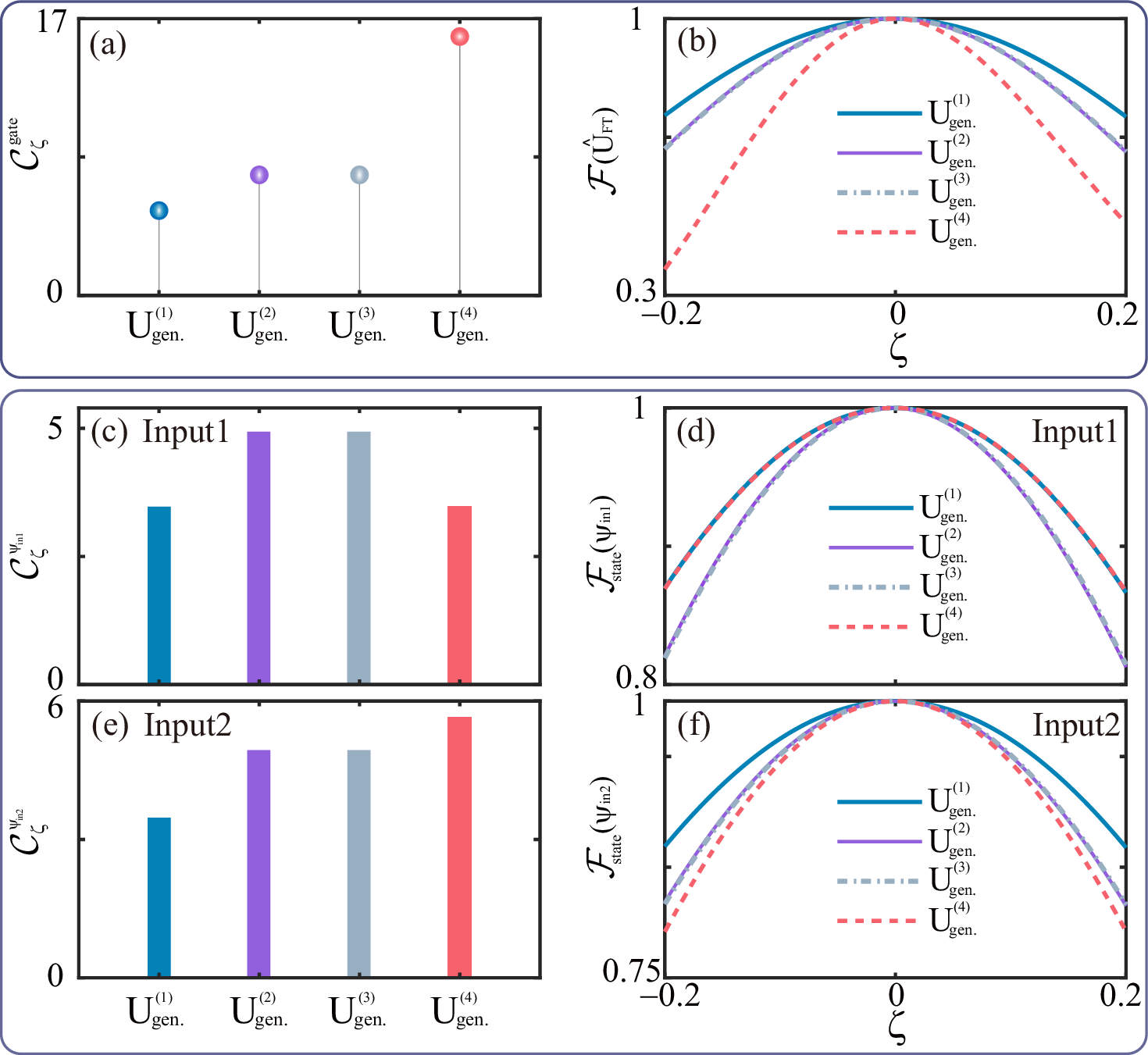} }\caption{(a) Analytical gate error coefficient $C_{\zeta}^{\rm gate}=d\,\mathrm{Tr}(\hat H_\alpha^2)-\mathrm{Tr}(\hat H_\alpha)^2$ [Eq.~(\ref{gateerr})] and (c,e) analytical state error coefficients $C_{\zeta}^{\psi_{\rm in}}=\langle \hat H_\alpha^2\rangle-\langle \hat H_\alpha\rangle^2$ [Eq.~(\ref{stateerr})] for input states $|\psi_1(t_0)\rangle=|0\rangle$ and $|\psi_2(t_0)\rangle=\frac{1}{\sqrt{3}}(|0\rangle+e^{i2\pi/3}|1\rangle+e^{-i2\pi/3}|2\rangle)$. (b) Average gate fidelity $\mathcal{F}(\hat U_{\rm FT})$ [Eq. (\ref{Figate})] and (d,f) corresponding state fidelities $\mathcal{F}_{\rm state}(\psi_{\rm in})$ [Eq. (\ref{Fidtate})] as functions of the phase error $\zeta$, defined by $\phi_{a,b,c} \rightarrow (1+\zeta)\phi_{a,b,c}$, for four decomposition sequences $\hat U_{\rm gen.}^{(1\text{-}4)}$. Other pulse parameters are the same as in Fig.~\ref{fig2}(f).} 
\label{fig6}
\end{figure}
We analyze the response of four SU(2) decomposition sequences to relative errors in the azimuthal angles, $\phi_{a,b,c}\rightarrow(1+\zeta)\phi_{a,b,c}$, representing phase perturbations in the control fields. Figure~\ref{fig6} shows the analytical error coefficients $\mathcal{C}_{\zeta}^{\rm gate}$ and
$\mathcal{C}_{\zeta}^{\psi_{\rm in}}$, together with the gate and state fidelities obtained from molecular simulations as functions
of the phase-error parameter $\zeta$. Figures~\ref{fig6}(a) and (b) indicate that the analytical gate-error coefficients vary across the four sequences, resulting in sequence-dependent changes in average gate fidelity as $\zeta$ changes. Unlike amplitude errors, phase errors remove the equivalence among SU(2) orderings at the gate level. The fourth decomposition, with the largest analytical error coefficient, is most sensitive to phase errors, while the first decomposition is most robust.

\begin{figure}[t]
\centering
\resizebox{0.5\textwidth}{!}{%
\includegraphics{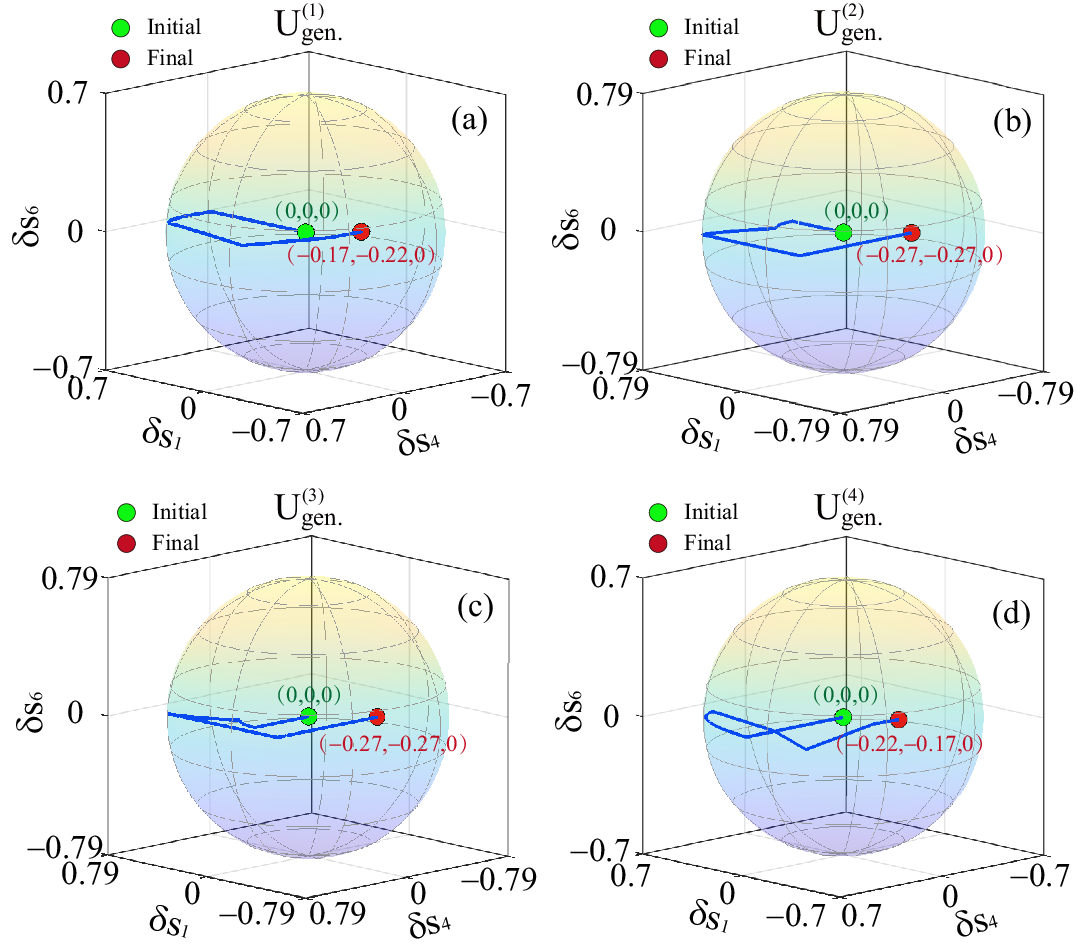} }\caption{The same simulations as in Fig. \ref{fig5}, but for phase errors with $\zeta=-0.2$, i.e., $\delta \mathbf{s}(t)=\mathbf{s}(\zeta=-0.2,t)-\mathbf{s}(0,t)$.} 
\label{fig7}
\end{figure}

This sequence dependence is also evident in state-to-state transformations. For the initial state $|\psi_1(t_0)\rangle$ [Figs.~\ref{fig6}(c) and (d)], the first and fourth decompositions share identical analytical error coefficients, while the second and third are more sensitive to phase perturbations, as shown by their state fidelities. For the coherent superposition input $|\psi_2(t_0)\rangle$
[Figs.~\ref{fig6}(e) and \ref{fig6}(f)], the second and third decompositions show similar responses, whereas the first and fourth exhibit different sensitivities, with the fourth being the most susceptible to phase errors. These findings show that phase errors create stronger sequence dependence than amplitude errors, impacting both overall gate fidelity and specific quantum-state transformations. The analytical error coefficients provide a quantitative basis for predicting and selecting decomposition orders to enhance robustness against phase noise.\\ \indent 
To clarify how phase errors accumulate during gate operations, we examine error dynamics using the generalized Bloch-vector (Gell-Mann) representation for a fixed phase perturbation ($\zeta=-0.2$) and the input state $|\psi_1(t_0)\rangle$. Figure~\ref{fig7} presents the time-dependent deviation $\delta s_k(t) = s_k(\zeta, t) - s_k(0, t)$ for each decomposition sequence. All four trajectories remain within the plane defined by $\delta s_6 = 0$, indicating that phase-error accumulation is confined to the $\lambda_1$ and $\lambda_4$ subspaces, which correspond to the real parts of the $|0\rangle\leftrightarrow|1\rangle$ and $|0\rangle\leftrightarrow|2\rangle$ coherences in the Gell-Mann basis.\\ \indent 
Unlike amplitude errors, phase errors do not propagate along the $\lambda_6$ direction and primarily affect coherence channels linked to the reference state $|0\rangle$. While all final deviations remain within this plane, the trajectory evolution depends on the decomposition sequence, indicating that phase-error propagation through multistep SU(2) operations is sequence specific. Therefore, phase-error accumulation in SU(3) space is governed by sequence-dependent dynamics.
\section{Conclusion and outlooks}\label{sec4}
We developed a theoretical framework for universal single-qutrit control in asymmetric-top molecules. The qutrit is encoded in three rotational eigenstates coupled by allowed single-photon transitions, with an auxiliary rotational state enabling independent phase control. Arbitrary qutrit gates are decomposed into three SU(2) rotations and a diagonal phase gate. By deriving the multilevel pulse-area theorem, we established a direct mapping between gate and control pulse parameters, allowing analytic design of microwave pulse sequences. To validate our method, we implemented the Walsh–Hadamard gate in 1,2-propanediol molecules. Simulations showed high-fidelity operation with negligible leakage to noncomputational states. Systematic analysis of amplitude and phase errors revealed that gate and state-fidelity sensitivities depend on the decomposition sequence, with error accumulation primarily along specific coherence channels. These results provide a theoretical basis for selecting decomposition sequences tailored to specific requirements, supporting the implementation of high-fidelity, robust single-qutrit gates in asymmetric-top molecular systems.

Experimental implementation of single-qutrit control requires two main capabilities. First, asymmetric-top molecules must be prepared and confined in their absolute ground state. Recent advances in molecular cooling have enabled progress toward individual molecule control \cite{2020Science_Doyle,2020PRL_Doyle,2022Nature_Doyle,2024PRL_Yan,2024NP_Ye,2024NP_Cornish}, such as with optical tweezers \cite{2019Science_Anderegg,2023Science_Bao,2023Science_Cheuk,2025Nature_Cornish}, though most techniques remain at the ensemble level. As methods advance, precise single-molecule manipulation is expected to become more feasible. Second, coherent manipulation of molecular rotational states relies on shaped multiple microwave pulse sequences with tunable amplitudes and phases. These sequences are achievable with current pulse-shaping and microwave-synthesis technologies \cite{2011OC_Yao,2014ACP,2021OL_Zeng,2024NC_Bao}, as demonstrated in atomic systems \cite{2015PRL_Saffman}. Applying these approaches to molecular platforms is feasible and could enable high-fidelity single-qutrit operations under realistic conditions.

Despite ongoing experimental challenges, particularly in achieving single-molecule control and precise state preparation, the theoretical framework established here provides a strong foundation for current and future experiments, as well as for extending these principles to more complex molecular qudit systems. Continued advances in molecular cooling, trapping, and coherent control are essential for realizing these protocols in quantum information platforms. Further research into error mitigation, optimized gate decomposition, and integration with scalable architectures will be critical for enabling robust, high-fidelity multilevel quantum logic in molecular systems.

 \begin{acknowledgments}
This work was supported by the National Natural Science Foundation of China under Grant No. 12274470.  The
        simulation was conducted using computing resources at the High Performance Computing Center of Central South University.
    \end{acknowledgments}

\appendix

\section{DERIVATION OF THE ELEMENTARY OPERATORS}\label{ApP_1}
We derive the elementary operators $\hat U_a$, $\hat U_b$, $\hat U_c$, and $\hat U(\eta,\chi)$ used in the main text. The first three represent SU(2) rotations driven by resonant couplings of the state pairs $\{|0\rangle,|1\rangle\}$, $\{|1\rangle,|2\rangle\}$,
and $\{|0\rangle,|2\rangle\}$, driven by the control fields $\vec{\mathcal E}_a(t)$, $\vec{\mathcal E}_b(t)$, and $\vec{\mathcal E}_c(t)$, respectively. Within the qutrit basis $\{|0\rangle,|1\rangle,|2\rangle\}$, the Hamiltonian in the interaction picture reads
\begin{equation}
\hat{H}_s(t)
=
-\begin{pmatrix}
0 & \mu_{01}\mathcal{E}_a(t)e^{-i\omega_{10}t} & \mu_{02}\mathcal{E}_c(t)e^{-i\omega_{20}t} \\
\mu_{10}\mathcal{E}_a(t)e^{i\omega_{10}t} & 0 & \mu_{12}\mathcal{E}_b(t)e^{-i\omega_{21}t} \\
\mu_{20}\mathcal{E}_c(t)e^{i\omega_{20}t} &  \mu_{21}\mathcal{E}_b(t)e^{i\omega_{21}t} & 0
\end{pmatrix},
\end{equation}
where $\mu_{ij}\mathcal E_n(t)=\langle i|\vec{\hat\mu}\cdot \vec{\mathcal E}_n(t)|j\rangle$.

Applying only $\vec{\mathcal{E}}_a(t)$, the first-order Magnus approximation yields
\begin{equation}\label{ua}
\hat{U}_a(\theta_a,\phi_a)=
\begin{pmatrix}
\cos\theta_a & i\sin\theta_a e^{i\phi_a} & 0 \\
i\sin\theta_a e^{-i\phi_a} & \cos\theta_a & 0 \\
0 & 0 & 1
\end{pmatrix},
\end{equation}
where $\theta_a$ and $\phi_a$ are determined by the complex pulse area $\theta_ae^{i\phi_a}=\mu_{01}\int \mathcal{E}_a(t')e^{-i\omega_{10}t'}dt'$.
This is an SU(2) rotation on $\{|0\rangle,|1\rangle\}$, leaving $|2\rangle$ unchanged.

Similarly, applying only $\vec{\mathcal E}_b(t)$ restricts the dynamics to $\{|1\rangle,|2\rangle\}$, giving
\begin{equation}\label{ub}
\hat{U}_b(\theta_b,\phi_b)=
\begin{pmatrix}
1 & 0 & 0 \\
0 & \cos\theta_b & i\sin\theta_b e^{i\phi_b} \\
0 & i\sin\theta_b e^{-i\phi_b} & \cos\theta_b
\end{pmatrix},
\end{equation}
where $\theta_be^{i\phi_b}=\mu_{12}\int \mathcal{E}_b(t')e^{-i\omega_{21}t'}dt'$.
Only $\vec{\mathcal{E}}_c(t)$ acts on $\{|0\rangle,|2\rangle\}$, producing
\begin{equation}\label{uc}
\hat{U}_c(\theta_c,\phi_c)=
\begin{pmatrix}
\cos\theta_c & 0 & i\sin\theta_c e^{i\phi_c} \\
0 & 1 & 0 \\
i\sin\theta_c e^{-i\phi_c} & 0 & \cos\theta_c
\end{pmatrix},
\end{equation}
where the complex pulse area satisfies $\theta_ce^{i\phi_c}=\mu_{02}\int \mathcal{E}_c(t')e^{-i\omega_{20}t'}dt'$.

To implement the phase gate, consider the auxiliary transition $|1\rangle \leftrightarrow |S\rangle$ driven by $\vec{\mathcal{E}}_P(t)$. In the basis $\{|1\rangle, |S\rangle\}$, the interaction-picture Hamiltonian is
\begin{equation}
    \hat{H}_{P}(t)=-\begin{pmatrix}
 0 &\mu_{1S}\mathcal{E}_P(t)e^{-i\omega_{1S}t} \\
 \mu_{1S}\mathcal{E}_P(t)e^{i\omega_{1S}t} &0
\end{pmatrix},
\end{equation}
where $\mu_{1S}\mathcal{E}_P(t)=\langle 1|\vec{\hat\mu}\cdot \vec{\mathcal{E}}_P(t)|S\rangle$. Assume $\vec{\mathcal{E}}_P(t)$ has two sequential subpulses $\vec{\mathcal{E}}_{P_1}(t)$, $\vec{\mathcal{E}}_{P_2}(t)$. The first-order Magnus approximation gives
\begin{equation}
\begin{aligned}
    \hat{U}_{P}&=\hat{U}_{P_2}(\theta_{P_2},\phi_{P_2})\hat{U}_{P_1}(\theta_{P_1},\phi_{P_1})\\
    &=\begin{pmatrix}
 \cos\theta_{P_2} &i\sin\theta_{P_2} e^{i\phi_{P_2}} \\
 i\sin\theta_{P_2} e^{-i\phi_{P_2}} & \cos\theta_{P_2} 
\end{pmatrix}\begin{pmatrix}
 \cos\theta_{P_1} &i\sin\theta_{P_1} e^{i\phi_{P_1}} \\
 i\sin\theta_{P_1} e^{-i\phi_{P_1}} & \cos\theta_{P_1} 
\end{pmatrix}.
\end{aligned} 
\end{equation}
Each pulse is a standard SU(2) rotation. For pulse areas $\theta_{P_1}=\theta_{P_2}=\theta_P=\pi/2$, the operator becomes diagonal:
\begin{equation}
   \hat{U}_{P}=\begin{pmatrix}
 e^{i\eta} &0 \\
 0 & e^{-i\eta} 
\end{pmatrix},
\end{equation}
with $\eta=\phi_{P_2}-\phi_{P_1}+\pi$. This sequence applies a pure phase shift to $|1\rangle$ without population transfer. Similarly, for $|2\rangle \leftrightarrow |S\rangle$ driven by $\vec{\mathcal{E}}_Q(t)$ (with two subpulses and pulse areas $\theta_{Q_1}=\theta_{Q_2}=\theta_Q=\pi/2$), the operator in $\{|S\rangle,|2\rangle\}$ is
\begin{equation}
    \hat{U}_{Q}=\begin{pmatrix}
 e^{-i\chi} &0 \\
 0 & e^{i\chi} 
\end{pmatrix},
\end{equation}
where $\chi=-\phi_{Q_2}+\phi_{Q_1}+\pi$. Combining both gives the phase gate in the qutrit basis:
\begin{equation}\label{uphase}
    \hat{U}(\eta,\chi)=\begin{pmatrix}
 1&0&0 \\
 0 & e^{i\eta} &0\\
 0 & 0& e^{i\chi}
\end{pmatrix}.
\end{equation}
Each operator represents an SU(2) rotation driven by a single resonant
transition. The phase gate is generated through a closed cyclic evolution
via the auxiliary state.

\section{DERIVATION OF AVERAGE GATE FIDELITY AND STATE FIDELITY}\label{ApP_2}
We derive the average gate and state fidelities in Eqs.~(\ref{gateerr}) and (\ref{stateerr}). For a gate $\hat U_{\mathrm{gen.}}(\alpha)$ with error parameter $\alpha\in\{\xi,\zeta\}$, define
\begin{equation}\label{Va}
\hat V(\alpha)
=
\hat U_{\mathrm{tar.}}^\dagger
\hat U_{\mathrm{gen.}}(\alpha),
\end{equation}
Here, $\hat U_{\mathrm{tar.}}=\hat U_{\mathrm{gen.}}(0)$ is the ideal gate and $\hat V(0)=\hat I$ (identity). Expanding $\hat U_{\mathrm{gen.}}(\alpha)$ [Eq. (\ref{Ualpha})] gives
\begin{equation}
\hat V(\alpha)=\hat I+\alpha \hat A+\frac{\alpha^2}{2}\hat B+\mathcal O(\alpha^3),
\end{equation}
where $\hat A=\hat U_{\mathrm{tar.}}^\dagger \frac{\partial \hat U_{\mathrm{gen.}}}{\partial \alpha}\big|_{\alpha=0}$ and $\hat B=\hat U_{\mathrm{tar.}}^\dagger \frac{\partial^2 \hat U_{\mathrm{gen.}}}{\partial\alpha^2}\big|_{\alpha=0}$. Unitarity of $\hat U_{\mathrm{gen.}}(\alpha)$ implies $\hat V(\alpha)$ is unitary. Expanding $\hat V^\dagger(\alpha)\hat V(\alpha)$ in $\alpha$ gives
\begin{equation}
\hat A^\dagger+\hat A=0,
\end{equation}
so $\hat A$ is anti-Hermitian. Define the Hermitian error generator
\begin{equation}
\hat H_\alpha=i\hat A=i\hat U_{\mathrm{tar.}}^{\dagger}\frac{\partial \hat U_{\mathrm{gen.}}}{\partial \alpha}\Bigg|_{\alpha=0}.
\end{equation}
At second order, expanding $\hat V^\dagger(\alpha)\hat V(\alpha)$ gives
\begin{equation}
\hat A^\dagger\hat A+\frac{1}{2}(\hat B^\dagger+\hat B)=0,
\end{equation}
which leads to
\begin{equation}
\hat B+\hat B^\dagger = -2\hat A^\dagger\hat A = -2\hat H_\alpha^2 .
\end{equation}
Thus,
\begin{equation}\label{Valpha}
\hat V(\alpha)=\hat I-i\alpha \hat H_\alpha+\frac{\alpha^2}{2}\hat B+\mathcal O(\alpha^3).
\end{equation}
The average gate fidelity is given by
\begin{equation}\label{FG}
\mathcal F_{\mathrm{gate}}(\alpha)=\frac{\left|\mathrm{Tr}\!\left[\hat U_{\mathrm{tar.}}^\dagger \hat U_{\mathrm{gen.}}(\alpha)\right]\right|^2+d}{d(d+1)}=\frac{|\mathrm{Tr}[\hat V(\alpha)]|^2+d}{d(d+1)} .
\end{equation}
Substituting Eq.~(\ref{Valpha}) into Eq.~(\ref{FG}), the trace term becomes
\begin{equation}
\mathrm{Tr}[\hat V(\alpha)]=d-i\alpha\,\mathrm{Tr}(\hat H_\alpha)+\frac{\alpha^2}{2}\mathrm{Tr}(\hat B)+\mathcal O(\alpha^3).
\end{equation}
Taking the trace and using $\hat H_\alpha^\dagger=\hat H_\alpha$ gives
\begin{equation}
\mathrm{Re}[\mathrm{Tr}(\hat B)]
=-\mathrm{Tr}(\hat H_\alpha^2).
\end{equation}
Therefore,
\begin{equation}
|\mathrm{Tr}\,\hat V(\alpha)|^2=d^2-\alpha^2\left[
d\,\mathrm{Tr}(\hat H_\alpha^2)-\mathrm{Tr}(\hat H_\alpha)^2
\right]+\mathcal O(\alpha^3),
\end{equation}
which yields
\begin{equation}
\mathcal F_{\mathrm{gate}}(\alpha)=1-\frac{\alpha^2}{d(d+1)}
\left[d\,\mathrm{Tr}(\hat H_\alpha^2)-\mathrm{Tr}(\hat H_\alpha)^2\right]+\mathcal O(\alpha^3).
\end{equation}

For input $|\psi_{\mathrm{in}}\rangle=|\psi(t_0)\rangle$, the ideal output is $|\psi_{\mathrm{tar.}}\rangle=\hat U_{\mathrm{tar.}}|\psi(t_0)\rangle$, and the realized output is $|\psi_{\mathrm{out}}\rangle=\hat U_{\mathrm{gen.}}(\alpha)|\psi(t_0)\rangle$.
The state fidelity is
\begin{equation}
\mathcal F_{\mathrm{state}}(\alpha)=\left|\langle\psi_{\mathrm{tar.}}|\psi_{\mathrm{out}}\rangle
\right|^2=\left|\langle \psi(t_0)|\hat V(\alpha)|\psi(t_0)\rangle\right|^2 .
\end{equation}
Expanding $\hat V(\alpha)$ yields
\begin{equation}
\langle \hat V(\alpha)\rangle=1-i\alpha \langle \hat H_\alpha\rangle+\frac{\alpha^2}{2}\langle \hat B\rangle+\mathcal O(\alpha^3),
\end{equation}
where \(\langle \hat O\rangle \equiv \langle \psi(t_0)|\hat O|\psi(t_0)\rangle\). Taking the modulus squared and using
$\mathrm{Re}\langle \hat B\rangle=-\langle \hat H_\alpha^2\rangle$, so
\begin{equation}
\mathcal F_{\mathrm{state}}(\alpha)=1-\alpha^2\left(\langle \hat H_\alpha^2\rangle-\langle \hat H_\alpha\rangle^2\right)+\mathcal O(\alpha^3).
\end{equation}
Thus, to leading order, state fidelity loss is set by the variance of $\hat H_\alpha$ in the input.

\section{GELL-MANN MATRICES FOR QUTRIT REPRESENTATION}\label{Appl}
The generalized Gell-Mann matrices for a qutrit are given by
$\{\lambda_k\}_{k=1}^8$:
\begin{equation}
\begin{aligned}
    &\lambda_1 = 
\begin{pmatrix}
0 & 1 & 0 \\
1 & 0 & 0 \\
0 & 0 & 0
\end{pmatrix}, \quad
\lambda_2 =
\begin{pmatrix}
0 & -i & 0 \\
i & 0 & 0 \\
0 & 0 & 0
\end{pmatrix}, \quad
\lambda_3 =
\begin{pmatrix}
1 & 0 & 0 \\
0 & -1 & 0 \\
0 & 0 & 0
\end{pmatrix},\\
&\lambda_4 =
\begin{pmatrix}
0 & 0 & 1 \\
0 & 0 & 0 \\
1 & 0 & 0
\end{pmatrix}, \quad
\lambda_5 =
\begin{pmatrix}
0 & 0 & -i \\
0 & 0 & 0 \\
i & 0 & 0
\end{pmatrix}, \quad
\lambda_6 =
\begin{pmatrix}
0 & 0 & 0 \\
0 & 0 & 1 \\
0 & 1 & 0
\end{pmatrix},\\
&\lambda_7 =
\begin{pmatrix}
0 & 0 & 0 \\
0 & 0 & -i \\
0 & i & 0
\end{pmatrix}, \quad
\lambda_8 = \frac{1}{\sqrt{3}}
\begin{pmatrix}
1 & 0 & 0 \\
0 & 1 & 0 \\
0 & 0 & -2
\end{pmatrix}.
\end{aligned}
\end{equation}
We can see that \(\lambda_1\) and \(\lambda_2\) represent coherences between \(|0\rangle\) and \(|1\rangle\); \(\lambda_4\) and \(\lambda_5\) between \(|0\rangle\) and \(|2\rangle\); and \(\lambda_6\) and \(\lambda_7\) between \(|1\rangle\) and \(|2\rangle\), whereas \(\lambda_3\) and \(\lambda_8\) indicate population differences among the three states.
\section{DECOMPOSITION OF THE WALSH-HADAMARD GATE}\label{para}
We now determine the decomposition parameters for the qutrit
Walsh-Hadamard gate and identify which sequences of the three pairwise SU(2) operations are admissible.
Its implementation requires solving the equation
\begin{equation}\label{FTeq}
\hat U(\eta,\chi)\,
\hat U_{m_1}(\theta_{m_1},\phi_{m_1})\,
\hat U_{m_2}(\theta_{m_2},\phi_{m_2})\,
\hat U_{m_3}(\theta_{m_3},\phi_{m_3})
=
\hat U_{\mathrm{FT}},
\end{equation}
for all six permutations of
\((m_1,m_2,m_3)\in\{(a,b,c),(a,c,b),(b,a,c),(b,c,a),(c,a,b),(c,b,a)\}\).

By substituting the matrix forms of \(\hat U_a\), \(\hat U_b\), and \(\hat U_c\),
and \(\hat U(\eta,\chi)\) [Eqs. (\ref{ua})-(\ref{uphase})] into Eq. (\ref{FTeq}) yields a set of algebraic equations
for the eight parameters
\((\theta_a,\phi_a,\theta_b,\phi_b,\theta_c,\phi_c,\eta,\chi)\). Among the six possible permutations, four produce real solutions:
\begin{equation}
\begin{aligned}
    \hat U_{\mathrm{FT}}
=
&\hat U(\eta,\chi) \hat U_c \hat U_a \hat U_b,
\quad
\hat U(\eta,\chi) \hat U_b \hat U_c \hat U_a,\\
&\hat U(\eta,\chi) \hat U_b \hat U_a \hat U_c,
\quad
\hat U(\eta,\chi) \hat U_a \hat U_c \hat U_b.
\end{aligned}
\end{equation}
The corresponding parameters are provided in Table \ref{tb1} of the main text.
To demonstrate that the remaining two permutations do not yield admissible real solutions, we first consider the ordering
\(\hat U(\eta,\chi) \hat U_c \hat U_b \hat U_a\).
The product of these four matrices is
\begin{equation}
\hat U(\eta,\chi) \hat U_c \hat U_b \hat U_a
=
\begin{pmatrix}
M_{11} & M_{12} & M_{13} \\
M_{21} & M_{22} & M_{23} \\
M_{31} & M_{32} & M_{33}
\end{pmatrix},
\end{equation}
with
\begin{equation}
\begin{aligned}
M_{21}&=i e^{i(\eta-\phi_a)} \sin\theta_a \cos\theta_b,\\
M_{22}&=e^{i\eta}\cos\theta_a\cos\theta_b,\\
M_{23}&=i e^{i(\eta+\phi_b)}\sin\theta_b,\\
M_{13}&=i e^{i\phi_c}\sin\theta_c\cos\theta_b.
\end{aligned}
\end{equation}
Since every matrix element of \(\hat U_{\mathrm{FT}}\) has modulus
\(1/\sqrt{3}\), comparison of the above entries with
\(\hat U_{\mathrm{FT}}\) gives
\begin{equation}
|\sin\theta_a\cos\theta_b|=|\cos\theta_a\cos\theta_b|=|\sin\theta_b|=
|\sin\theta_c\cos\theta_b|=\frac{1}{\sqrt{3}}.
\end{equation}
This implies that
\begin{equation}
\theta_a=\theta_c=\frac{\pi}{4},
\quad
\theta_b=\arcsin\!\frac{1}{\sqrt{3}}.
\label{A:angles1}
\end{equation}
The phases are determined by the entries
\(M_{22}=e^{i2\pi/3}/\sqrt{3}\),
\(M_{21}=1/\sqrt{3}\),
\(M_{23}=e^{i4\pi/3}/\sqrt{3}\),
and
\(M_{13}=1/\sqrt{3}\), which gives
\begin{equation}
\eta=\frac{2\pi}{3},
\quad
\phi_a=\frac{7\pi}{6},
\quad
\phi_b=\frac{\pi}{6},
\quad
\phi_c=\frac{3\pi}{2}.
\label{A:phases1}
\end{equation}
Substituting Eqs.~(\ref{A:angles1}) and (\ref{A:phases1}) into the
remaining matrix element \(M_{12}\) gives
\begin{equation}
M_{12}
=
\frac{1}{4}+\frac{\sqrt{3}}{12}
+
i\left(\frac{1}{4}-\frac{\sqrt{3}}{4}\right),
\end{equation}
which does not correspond to the target value
$(\hat{U}_{\mathrm{FT}})_{12}=1/\sqrt{3}$.
Therefore, the decomposition
\(\hat U_p \hat U_c \hat U_b \hat U_a\) is inconsistent and does not
yield a valid solution.

Next, we consider the ordering
\(\hat U_p \hat U_a \hat U_b \hat U_c\).
In this case, the relevant matrix elements are
\begin{equation}
\begin{aligned}
 M_{12}&=i e^{i\phi_a}\sin\theta_a\cos\theta_b,\\
M_{22}&=e^{i\eta}\cos\theta_a\cos\theta_b,\\
M_{32}&=i e^{i(\chi-\phi_b)}\sin\theta_b,\\
M_{31}&=i e^{i(\chi-\phi_c)}\sin\theta_c\cos\theta_b.
\end{aligned}
\end{equation}
Matching their moduli to those of \(\hat U_{\mathrm{FT}}\) again gives
\begin{equation}
\theta_a=\theta_c=\frac{\pi}{4},
\quad
\theta_b=\arcsin\!\frac{1}{\sqrt{3}},
\quad
\cos\theta_b=\sqrt{\frac{2}{3}}.
\label{A:angles2}
\end{equation}
The phases are determined by \(M_{12}=1/\sqrt{3}\), \(M_{22}=e^{i2\pi/3}/\sqrt{3}\), \(M_{32}=e^{i4\pi/3}/\sqrt{3}\), \(M_{31}=1/\sqrt{3}\), and \(M_{33}=e^{i2\pi/3}/\sqrt{3}\), resulting in
\begin{equation}\label{A:phases2}
\phi_a=\frac{3\pi}{2},\quad \eta=\frac{2\pi}{3}, \quad
\chi=\frac{2\pi}{3},\quad\phi_b=\frac{11\pi}{6},\quad\phi_c=\frac{7\pi}{6}.
\end{equation}
Substituting Eqs.~(\ref{A:angles2}) and (\ref{A:phases2}) into the matrix element \(M_{11}\) yields
\begin{equation}
M_{11}=\frac{1}{2}+\frac{1}{4\sqrt{3}}-\frac{i}{4},
\end{equation}
which differs from the target value \((\hat U_{\mathrm{FT}})_{11}=1/\sqrt{3}\). Therefore, the decomposition \(\hat U_p \hat U_a \hat U_b \hat U_c\) is also inconsistent. This analysis confirms that only four of the six possible decomposition sequences yield valid real solutions for the qutrit Walsh-Hadamard gate.


%

\end{document}